\documentclass[aps,prl,twocolumn,superscriptaddress,showpacs,preprintnumbers,amsmath,amssymb]{revtex4}

\usepackage{graphicx} 
\usepackage{dcolumn}  
\usepackage{color}  

\graphicspath{{ps}}

\begin{document}


\title{ \quad\\[1.0cm] Observation of {\boldmath $B^0 \to 
D^{*-} \tau^+
  \nu_{\tau}$} decay at Belle}

\affiliation{Budker Institute of Nuclear Physics, Novosibirsk}
\affiliation{Chiba University, Chiba}
\affiliation{University of Cincinnati, Cincinnati, Ohio 45221}
\affiliation{Department of Physics, Fu Jen Catholic University, Taipei}
\affiliation{The Graduate University for Advanced Studies, Hayama}
\affiliation{Hanyang University, Seoul}
\affiliation{University of Hawaii, Honolulu, Hawaii 96822}
\affiliation{High Energy Accelerator Research Organization (KEK), Tsukuba}
\affiliation{Hiroshima Institute of Technology, Hiroshima}
\affiliation{Institute of High Energy Physics, Chinese Academy of Sciences, Beijing}
\affiliation{Institute of High Energy Physics, Vienna}
\affiliation{Institute of High Energy Physics, Protvino}
\affiliation{Institute for Theoretical and Experimental Physics, Moscow}
\affiliation{J. Stefan Institute, Ljubljana}
\affiliation{Kanagawa University, Yokohama}
\affiliation{Korea University, Seoul}
\affiliation{Kyungpook National University, Taegu}
\affiliation{Swiss Federal Institute of Technology of Lausanne, EPFL, Lausanne}
\affiliation{University of Ljubljana, Ljubljana}
\affiliation{University of Maribor, Maribor}
\affiliation{University of Melbourne, School of Physics, Victoria 3010}
\affiliation{Nagoya University, Nagoya}
\affiliation{Nara Women's University, Nara}
\affiliation{National Central University, Chung-li}
\affiliation{National United University, Miao Li}
\affiliation{Department of Physics, National Taiwan University, Taipei}
\affiliation{H. Niewodniczanski Institute of Nuclear Physics, Krakow}
\affiliation{Nippon Dental University, Niigata}
\affiliation{Niigata University, Niigata}
\affiliation{University of Nova Gorica, Nova Gorica}
\affiliation{Osaka City University, Osaka}
\affiliation{Osaka University, Osaka}
\affiliation{Panjab University, Chandigarh}
\affiliation{University of Science and Technology of China, Hefei}
\affiliation{Seoul National University, Seoul}
\affiliation{Sungkyunkwan University, Suwon}
\affiliation{University of Sydney, Sydney, New South Wales}
\affiliation{Toho University, Funabashi}
\affiliation{Tohoku Gakuin University, Tagajo}
\affiliation{Tohoku University, Sendai}
\affiliation{Department of Physics, University of Tokyo, Tokyo}
\affiliation{Tokyo Institute of Technology, Tokyo}
\affiliation{Tokyo Metropolitan University, Tokyo}
\affiliation{Tokyo University of Agriculture and Technology, Tokyo}
\affiliation{Virginia Polytechnic Institute and State University, Blacksburg, Virginia 24061}
\affiliation{Yonsei University, Seoul}
  \author{A.~Matyja}\affiliation{H. Niewodniczanski Institute of Nuclear Physics, Krakow} 
  \author{M.~Rozanska}\affiliation{H. Niewodniczanski Institute of Nuclear Physics, Krakow} 
  \author{I.~Adachi}\affiliation{High Energy Accelerator Research Organization (KEK), Tsukuba} 
  \author{H.~Aihara}\affiliation{Department of Physics, University of Tokyo, Tokyo} 
  \author{V.~Aulchenko}\affiliation{Budker Institute of Nuclear Physics, Novosibirsk} 
  \author{T.~Aushev}\affiliation{Swiss Federal Institute of Technology of Lausanne, EPFL, Lausanne}\affiliation{Institute for Theoretical and Experimental Physics, Moscow} 
  \author{S.~Bahinipati}\affiliation{University of Cincinnati, Cincinnati, Ohio 45221} 
  \author{A.~M.~Bakich}\affiliation{University of Sydney, Sydney, New South Wales} 
  \author{V.~Balagura}\affiliation{Institute for Theoretical and Experimental Physics, Moscow} 
  \author{E.~Barberio}\affiliation{University of Melbourne, School of Physics, Victoria 3010} 
% ...
\collaboration{The Belle Collaboration}
\noaffiliation

\begin{abstract}
We report an observation of the decay 
$B^0\to D^{*-} \tau^+ \nu_{\tau}$ in a 
data sample 
containing $535\times10^6$ $B\bar{B}$ pairs 
collected with the Belle detector at the KEKB 
asymmetric-energy $e^+e^-$ 
collider. We find a signal with a 
significance of 
5.2$\sigma$ and
measure the branching fraction  $\mathcal{B}(B^0\to D^{*-} \tau ^+ 
\nu_{\tau})=(2.02 ^{+0.40}_{-0.37} (stat) \pm 0.37 (syst)) \% $.  This 
is the first 
observation of an 
exclusive $B$ decay with  a $b \to  c \tau \nu_{\tau}$      
transition.
\end{abstract}


\maketitle

\tighten

{\renewcommand{\thefootnote}{\fnsymbol{footnote}}}
\setcounter{footnote}{0}


$B$ meson decays with $b \to  c \tau \nu_{\tau}$ transitions
can provide important constraints on the Standard 
Model (SM) and its extensions. 
Due to the large mass of the lepton in the final state  
these decays are sensitive 
probes of models with extended Higgs sectors \cite{Itoh} 
and provide observables sensitive to new physics, 
such as polarizations, 
which cannot be accessed in 
other semileptonic decays. 

Multiple neutrinos in the final states make the search for 
semi-tauonic $B$ decays very challenging  and 
hence there is little experimental information 
about these processes. So far, results are limited to inclusive and 
semi-inclusive measurements by LEP experiments \cite{lep} 
which measure 
an average 
branching fraction of 
$\mathcal{B}(b \to  \tau \nu_{\tau} X)=(2.48\pm 
0.26)\%$ 
\cite{PDG}.  
SM calculations predict branching fractions 
for $B\to \bar{D}^* \tau ^+ \nu_{\tau}$ 
around 1.4\% with uncertainties arising mainly from
assumptions about form-factors \cite{hwang}. 

In this paper we present 
the first observation of
$B^0\to D^{*-} \tau ^+ \nu_{\tau}$ \cite{CC}
decay using a data sample 
containing $535\times10^6$ $B\bar{B}$ pairs that 
were collected with 
the Belle detector at the KEKB asymmetric-energy $e^+e^-$ (3.5 on 8 GeV) 
collider \cite{KEKB} operating at the $\Upsilon(4S)$ resonance 
($\sqrt{s}=10.58$ GeV). 
The Belle 
detector is a large-solid-angle magnetic spectrometer consisting of a 
silicon vertex detector, a 50-layer central drift chamber, a 
system of aerogel Cherenkov counters, time-of-flight scintillation 
counters and an electromagnetic calorimeter (ECL) comprised of CsI(Tl) 
crystals located inside a superconducting solenoid coil that 
provides a 1.5 T magnetic field. An iron flux-return located outside the 
coil is instrumented to identify  $K_L^0$ mesons 
and muons. 
A detailed description 
of the detector can be found in Ref.~\cite{Belle}. 
We use Monte Carlo (MC) simulations to estimate signal
efficiencies and background contributions.
Large samples of the signal 
$B^0\to D^{*-} \tau ^+ \nu_{\tau}$ decays are generated with the EvtGen 
package \cite{evtgen} 
using the ISGW2 model \cite{isgw2}. Radiative effects are modeled by the PHOTOS 
code \cite{photos}. 
MC samples equivalent to  
about twice the accumulated data are 
used to evaluate the background from 
$B\bar{B}$ 
and continuum $q\bar{q}$ ($q=u,d,s,c$) events.  
       
$B$ decays to multi-neutrino final states 
can be observed at B-factories via the recoil 
of the accompanying 
$B$ meson ($B_{\rm tag}$) \cite{Ikado}.
Reconstruction of the 
$B_{\rm tag}$ strongly 
suppresses the combinatorial and continuum backgrounds and provides kinematical 
constraints on the signal meson ($B_{\rm sig}$).
In this study we take advantage of 
the clean signature provided by the $D^*$ meson occurring on 
the signal side 
and reconstruct the $B_{\rm tag}$ ``inclusively'' from all the 
particles that 
remain after selecting candidates for $B_{\rm sig}$ daughters. We apply 
the analysis to $B_{\rm sig}$ decay chains that combine 
a high reconstruction efficiency with a low background level.
The  $D^{*-}$ mesons are reconstructed in the $D^{*-}\to\bar{D}^0 \pi^-$ 
decay 
channel. The $\bar{D}^0$'s are reconstructed in the $K^+\pi^-$ and 
$K^+\pi^-\pi^0$ final states.
The $\tau^+ \to e^+\nu_e\bar{\nu}_{\tau}$  and $\tau^+ \to 
\pi^+\bar{\nu}_{\tau}$ modes are used to reconstruct 
$\tau$ lepton candidates.
We do not include the $\tau^+\to \mu^+ 
\nu_{\mu}\bar{\nu}_{\tau}$ mode because in the 
relevant momentum range the muon identification is inefficient. 
The $\tau^+\to \pi^+\bar{\nu}_{\tau}$ channel has higher combinatorial 
background 
than the purely leptonic mode, but the single neutrino in $\tau$ decay 
provides better kinematical constraints. 
For this mode we analyze only the 
$\bar{D}^0 \to K^+ \pi^-$ decay.

We select charged tracks with impact parameters that are 
consistent with an origin at the beam spot, and having momenta above
50 MeV/$c$ in the laboratory frame.
Muons, electrons, charged pions, kaons 
and (anti)protons are identified using information from particle 
identification subsystems.  The electrons from signal decays are 
selected with an efficiency greater than 90\% and a misidentification 
rate 
below 0.2\%. The momenta of particles identified as electrons  are 
corrected for bremsstrahlung by adding photons within a 50 mrad cone 
along the trajectory.
The  $\pi^0$ candidates are reconstructed from photon pairs having 
invariant mass in the range 118 MeV/$c^2<M_{\gamma\gamma}<$150 
MeV/$c^2$. 
From candidates 
that share a common $\gamma$, we select the $\pi^ 0$ with the 
smallest  $\chi^2$ value from a mass-constrained fit. 
To reduce the combinatorial background, we 
require photons from the $\pi^ 0$ to have energies 
above 60 MeV -120 MeV, 
depending on the photon's polar angle.
Photons  that do not come from a 
$\pi^0$ and exceed a polar-angle dependent energy 
threshold (100 MeV - 
200 MeV) are included in the $B_{\rm tag}$ reconstruction. 

We reconstruct the signal decay by selecting 
combinations of a
$D^{*-}$ meson 
and an electron or a pion candidate
with opposite charge. 
We accept $\bar{D}^0$ 
candidates with invariant masses in a 5$\sigma$  window around the 
nominal PDG \cite{PDG} value.
$D^{*-}$ candidates  are accepted if the mass difference  
$M_{D^*}-M_{D^0}$ is in a 3$\sigma$  
window around the PDG value. In order to reduce background from 
incorrectly reconstructed 
tracks, we impose tighter 
impact parameter requirements 
on the $e$ and $\pi$  candidates from $\tau$  decay. 
        
\begin{figure}[htb]
\includegraphics[width=0.22\textwidth, angle=0]{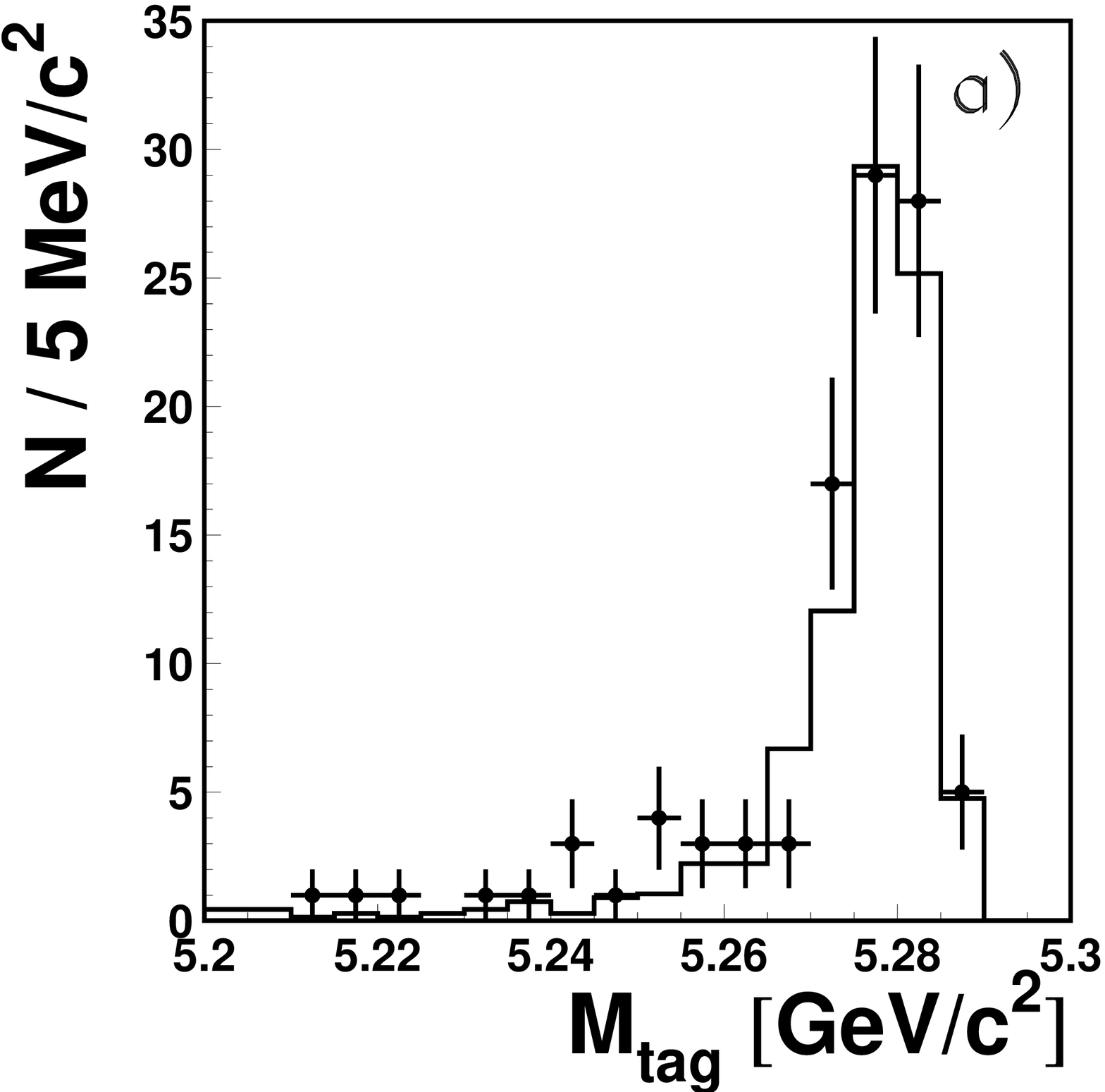}
\includegraphics[width=0.22\textwidth, angle=0]{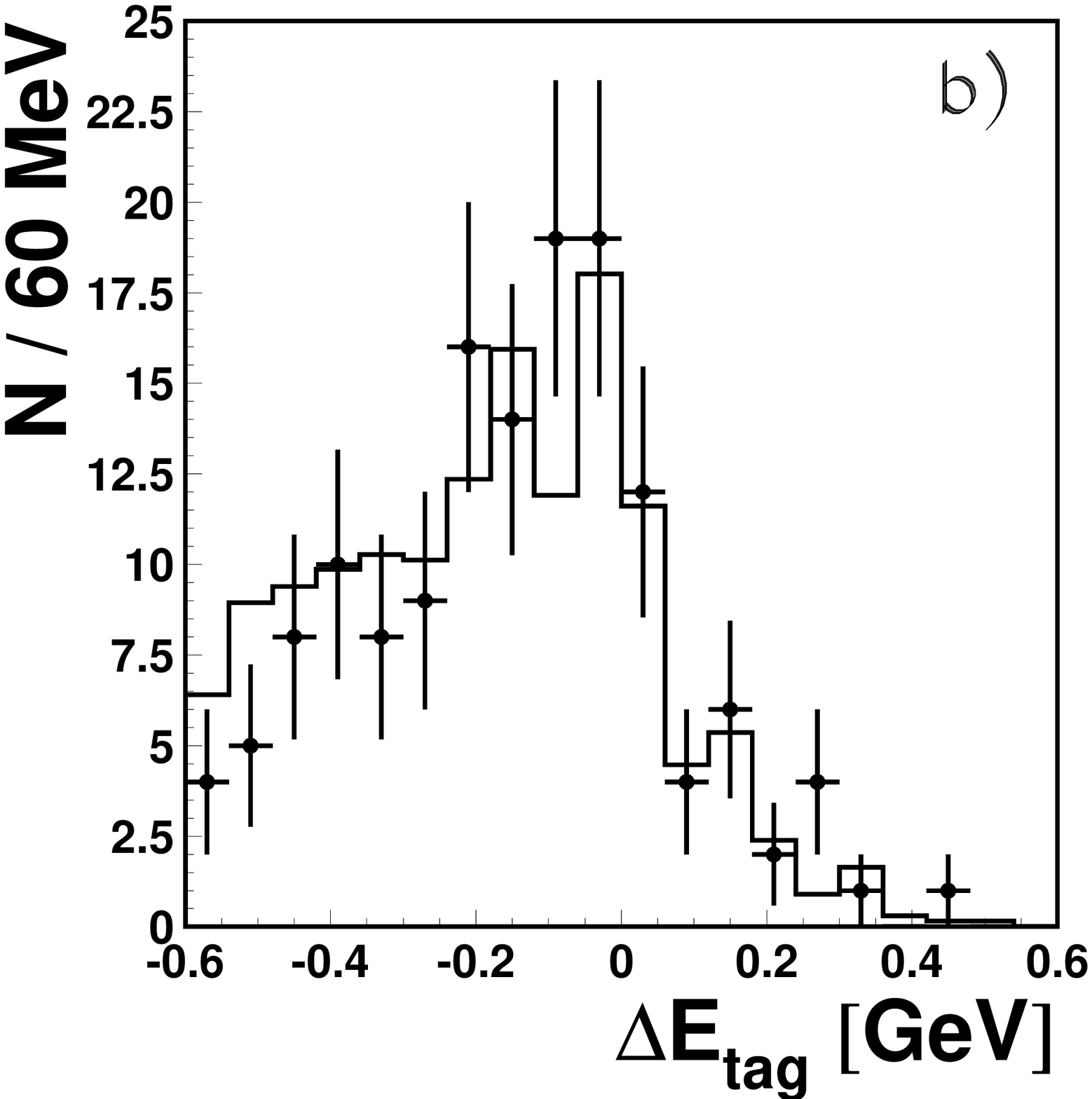}
\caption{$M_{\rm tag}$ and $\Delta E_{\rm tag}$ distributions 
for $B^0 \to 
D^{*-} 
\pi^+$ control sample from data (points with error bars) and MC 
(histograms).}
\label{pic-dstpi}
\end{figure}

Once a $B_{\rm sig}$ candidate is found, the remaining particles are 
used to reconstruct the $B_{\rm tag}$ decay.  The consistency of a 
$B_{\rm tag}$ candidate 
with a $B$-meson decay is checked using the beam-energy constrained mass and 
the energy difference variables:
$M_{\rm tag} = \sqrt{E^2_{\rm beam} - {\bf p}^2_{\rm tag}},~~
{\bf p}_{\rm tag} \equiv \sum_i {\bf p}_i$, and
$\Delta E_{\rm tag} = E_{\rm tag} - E_{\rm beam}, ~~ E_{\rm tag} 
\equiv 
\sum_i E_i$,
where $E_{\rm beam}$ 
is the beam energy and ${\bf p}_i$ and $E_i$ 
denote the momentum vector and energy of the $i$'th particle
in the $\Upsilon(4S)$ rest frame.
The summation is over all particles that are not assigned 
to $B_{\rm sig}$ and 
satisfy the selection criteria described above.  
We require that events have at least one 
($D^{*-}e^+/\pi^+$) pair and that 
$M_{\rm tag}$ and  $E_{\rm tag}$ satisfy $M_{\rm tag}>$5.2 GeV/$c^2$
and $|\Delta E_{\rm tag}|<$0.6 GeV. 
To improve 
the quality of the $B_{\rm tag}$ reconstruction, we impose the following 
requirements: zero total event charge, no $\mu^\pm$      
and no additional $e^\pm$ in  the event, zero net proton/antiproton 
number, residual energy in the ECL 
({\it i.e.} the sum of energies of clusters that do not 
fulfill the requirements imposed 
on photons) less than 0.35 GeV and number of neutral particles
on the tagging side $N_{\pi^0}+N_{\gamma}<$5.  
These criteria, which we refer to as 
``the $B_{\rm tag}$-selection'', reject events in 
which some particles were undetected and 
suppress events with a large number of spurious showers.  
In order to 
validate the $B_{\rm tag}$ simulation and reconstruction, we use a 
control sample of events, where the $B_{\rm sig}$ decays to 
$D^{*-}\pi^+$ (followed by 
$D^{*-}\to\bar{D}^0\pi^-$, $\bar{D}^0\to K^+\pi^-$) which allows us 
to select a $B\bar{B}$ sample with a purity of 96\% 
and with 
$B_{\rm sig}$ and $B_{\rm tag}$ daughters properly assigned to the 
parent particles.
Figure 1 shows the 
$M_{\rm tag}$ and  $\Delta E_{\rm tag}$ distributions of the control sample 
for data and the MC simulation scaled to the integrated luminosity in 
data.  
The events satisfy the $B_{\rm tag}$-selection criteria and 
are in the $-0.25$ GeV$<\Delta E_{\rm tag}<$ 0.05 GeV 
(for Fig.~\ref{pic-dstpi}(a)) and 
$M_{\rm tag}>$5.27 GeV/$c^2$ (for Fig.~\ref{pic-dstpi}(b)) windows. 
The good 
agreement of the shapes and of the absolute normalization demonstrates 
the validity of the MC-simulations for $B_{\rm tag}$ decays. 
Based on this 
study we constrain all further analysis to the region  
$-0.25$ GeV$<\Delta E_{\rm tag}<$0.05 GeV.
With this requirement 
about 80\% of the events are contained in the range
$M_{\rm tag}>5.27$ GeV/$c^2$.
    
\begin{figure}[htb]
\includegraphics[width=0.23\textwidth, angle=0]{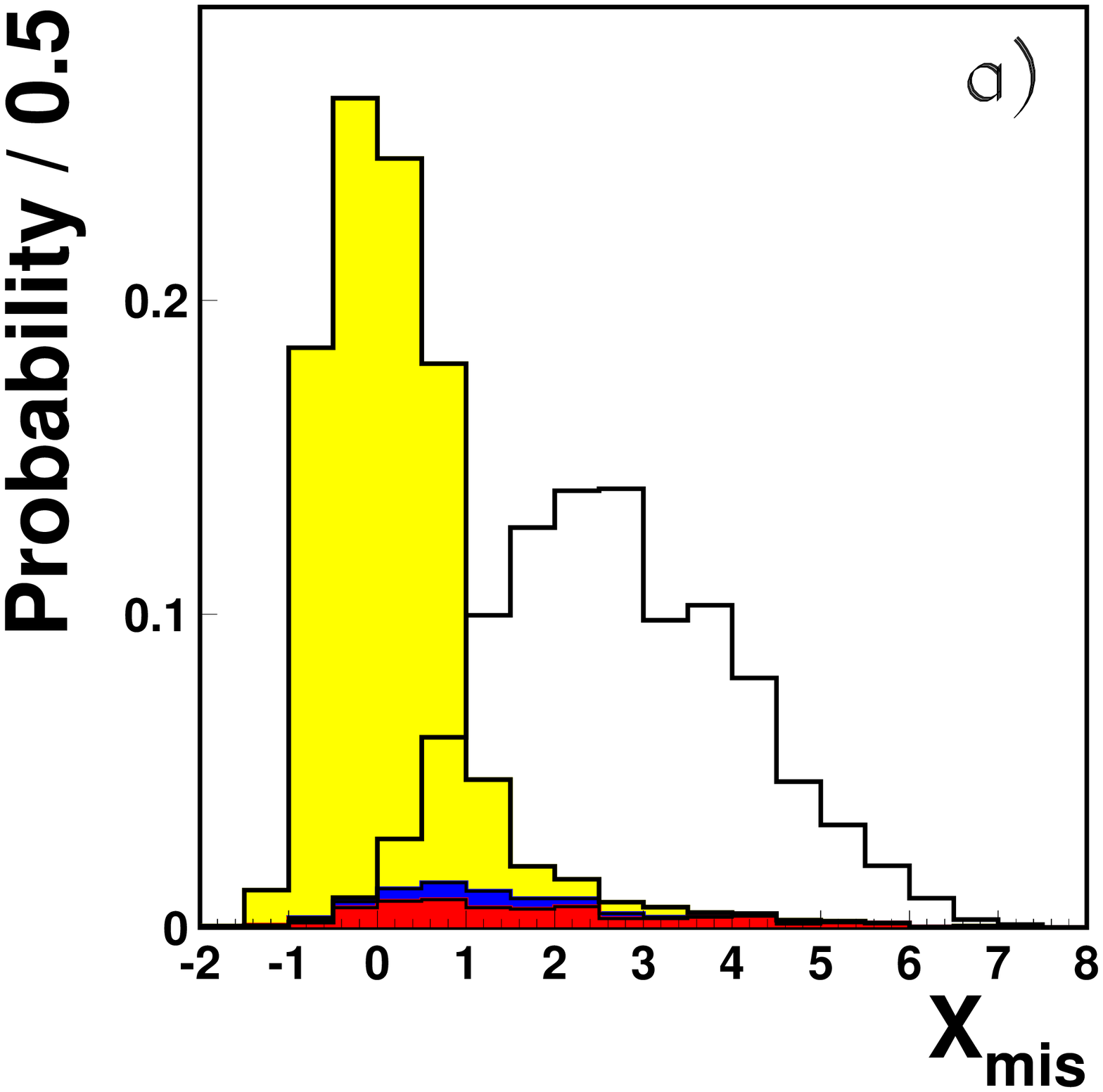}
\includegraphics[width=0.23\textwidth, angle=0]{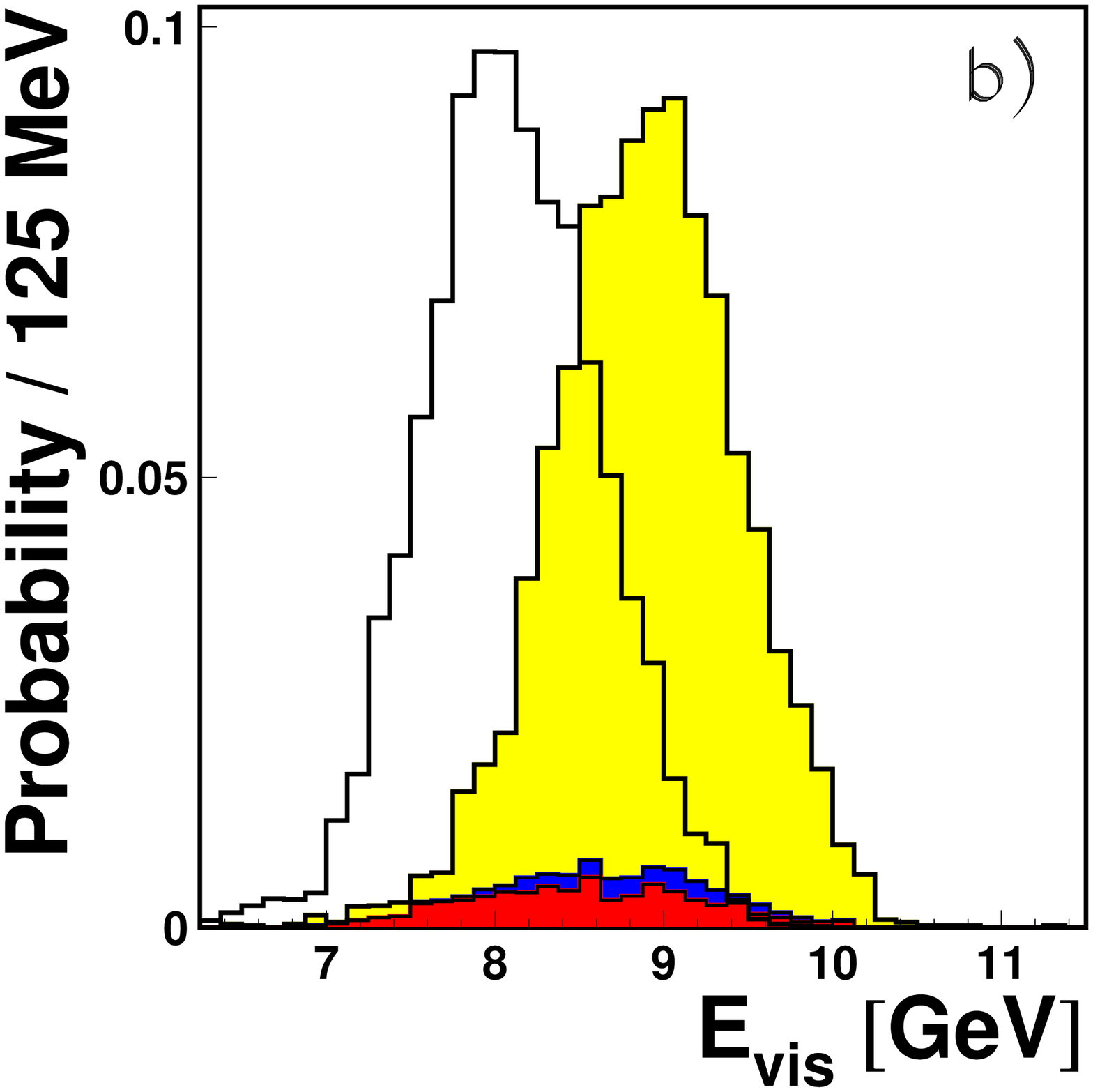}
\caption{
$X_{\rm mis}$ and $E_{\rm vis}$ distributions (normalized to unity)
after the $B_{\rm tag}$-selection 
for signal (blank) and 
background (shaded)
for the $\tau \to e\nu\nu$ mode
in the region $M_{\rm tag}>5.27$ GeV/$c^2$.  
The background
components, from top to bottom: 
$B^0\to D^{*-}e^+\nu_e$,
$B\to D^{**}e^+\nu_e$, and 
other $B$ decays.
The contribution from $q\bar{q}$-continuum 
is negligible.}  
\label{pic-cuts}
\end{figure}

The procedure described above, when applied to events with  
($D^{*-}e^+$) pairs
selects a relatively clean sample of semileptonic $B$ decays with the 
dominant non-signal contribution from the $B^0\to 
D^{*-}e^+\nu_{e}$ mode.  
Combinatorial background from hadronic $B$-decays dominates in 
the $\tau^+ \to \pi^{+}\bar{\nu}_{\tau}$ mode.
The background suppression exploits observables that characterize the 
signal decay: missing energy $E_{\rm mis} = 
E_{\rm beam}-E_{D^*}-E_{e/\pi}$; 
visible energy $E_{\rm vis}$, {\it i.e.} 
the sum of the energies of all particles in the event; 
the square of missing mass 
$M_{\rm mis}^2 = E_{\rm mis}^2 - 
({\bf p}_{\rm sig} - {\bf p}_{D^*} -
{\bf p}_{e/\pi})^2$ and the effective mass of the 
($\tau \nu_{\tau}$) pair, $M_{\rm W}^2 = (E_{\rm beam} - 
E_{D^*})^2 - ({\bf p}_{\rm sig} - 
{\bf p}_{D^*})^2$ where ${\bf p}_{\rm sig} = -{\bf p}_{\rm tag}$.
The most powerful variable for separating signal and background is 
obtained by combining $E_{\rm mis}$ and ($D^{*} 
e/\pi$) pair momentum: $ 
X_{\rm mis} \equiv (E_{\rm mis} - |{\bf p}_{D^*} + 
{\bf p}_{e/\pi}|)/
\sqrt{E_{\rm beam}^2 -m_{B^0}^2}$ where $m_{B^0}$ is the $B^0$ mass.
The $X_{\rm mis}$ variable 
is closely related to
the missing mass in the $B_{\rm sig}$ decay and does not depend on 
$B_{\rm tag}$ reconstruction. 
It lies in the  range $[-1, 1]$ for 
events with zero missing mass ({\it e.g.} with a single neutrino)  and 
takes larger values if there are multiple neutrinos. 
The MC distributions of $X_{\rm mis}$ and $E_{\rm vis}$ 
for signal and background events after 
$B_{\rm tag}$-selection
for the $\tau \to e \nu \nu$  mode
are shown in Fig.~\ref{pic-cuts}.  
The relative normalizations of the main background 
categories, $B^0\to D^{*-}e^+\nu_e$, $B\to D^{**}e^+\nu_e$, 
other $B$ decays and $q\bar{q}$ continuum, are determined from the 
data using looser selection criteria 
and verified using the sideband 
regions of the data sample that passed the 
final signal selection.   

We optimize selection criteria using MC samples for signal
and backgrounds, 
separately for decay chains 
with $\tau \to e \nu \nu$ and with $\tau \to \pi \nu$.  
In the first case we require  $X_{\rm mis}>$2.75, 1.9 
GeV$<E_{\rm mis}<$2.6 GeV and  
$E_{\rm vis}<$8.3 GeV.
We also reject events with a small difference between
$M_{\rm W}^2$ and $M_{\rm mis}^2$ to
suppress background from hadronic $B$ decays where
a genuine $D^*$ meson is combined with a soft 
secondary $e^\pm$. 
Decays in the $\tau \to \pi \nu$ mode are 
selected by requiring $X_{\rm mis}>$1.5, 
$M_{\rm W}^2-M_{\rm mis}^2-m_{\tau}^2+m_{\pi}^2>$0  
($m_{\tau}$ and $m_{\pi}$ denote the masses of the $\tau$ and charged $\pi$,
respectively), $E_{\rm vis}<$8.3 GeV,
the energy of the $\pi^+$ from the ($D^{*-}\pi^+$) pair greater than 0.6 
GeV, no $K^0_L$ in the event and less than four tracks that do not satisfy
the requirements imposed on the impact parameters.
The second requirement is equivalent to the condition 
$|\cos\theta_{\nu_1\nu_2}|<$1, 
where   $\theta_{\nu_1\nu_2}$ 
denotes the angle between the two neutrinos in the 
($\tau^+\nu_{\tau}$) rest frame.
The last three criteria reduce combinatorial background from 
low momentum pions 
and background from hadronic $B\to D^{*-}K_L^0+X$ and  
$B\to D^{*-}n\bar{n}+X$ decays.  
The above 
requirements result in flat $M_{\rm tag}$ distributions for 
most background 
components, while the signal distribution remains unchanged. This allows 
us to use the $M_{\rm tag}$ variable to extract the signal.

The $M_{\rm tag}$ distribution of the signal is described using 
a Crystal Ball (CB) lineshape function \cite{CB}. 
The shape parameters of the CB-function are determined from unbinned 
maximum likelihood fits to the combined MC signal samples. 
All the fits are performed in the range $M_{\rm tag}>5.2$ 
GeV/$c^2$.
The backgrounds are modeled as the sum of 
a combinatorial component using a parameterization introduced by ARGUS
(ARGUS-function) 
\cite{ARGUS} and a 
peaking background described by the CB-function with shape 
parameters fixed from the fit to the signal MC. 
 The main source of the peaking 
background is 
the semileptonic decay $B^0 \to D^{*-}e^+\nu_e$.
Cross-feed events from signal decays followed by $\tau$  decays 
to other modes are negligible in the 
$\tau \to e \nu \nu$ mode, but give significant 
contributions to the $\tau \to \pi \nu$ mode.
About half of the cross-feed comes from $\tau \to \rho \nu$ 
decay. 
We parameterize 
the $M_{\rm tag}$ distribution of cross-feed events as a sum 
of CB and ARGUS functions with shape parameters fixed from fits 
to the signal and combinatorial background as described above. The 
component described by 
the CB-function is treated as a part of the signal.
The efficiencies of signal reconstruction and  
the expected combinatorial and peaking backgrounds 
are given 
in Table~\ref{table-yields}. 

\begin{figure}[htb]
\includegraphics[width=0.35\textwidth, angle=0]{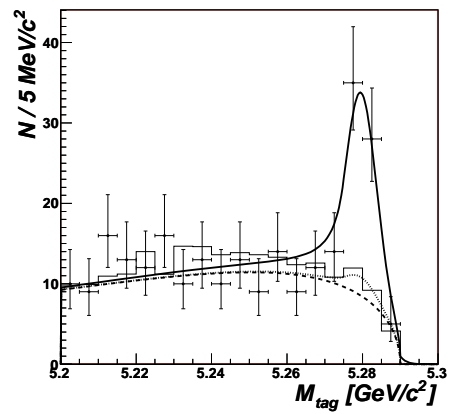}
\caption{
$M_{\rm tag}$ distribution for the combined data sample. The histogram 
represents expected background scaled to the data luminosity.
The solid curve shows the result of the fit. The dotted and dashed curves 
indicate respectively the fitted background and the 
component described by the ARGUS-function.} 
\label{pic-mtag}
\end{figure}

The selection criteria established in the MC studies are applied to the 
data. The resulting $M_{\rm tag}$ distribution for 
data in all three decay chains 
is shown in Fig.~\ref{pic-mtag}. 
The overlaid histogram represents the expected background, scaled to the data 
luminosity. A clear excess over  
background can be observed. 

We extract signal yields by fitting the $M_{\rm tag}$ distributions to 
the sum of the expected signal and background distributions using the 
following likelihood function:
\begin{eqnarray}
\mathcal{L} = 
e^{-(N_s+N_{p}+N_b)}\prod^N_{i=1}[(N_s+N_p)P_s(x_i)+N_b 
P_b(x_i)],
\end{eqnarray}
where $x_i$ is the $M_{\rm tag}$ in the $i$'th event and $N$ is 
the total number of events in the data.
$P_s$ ($P_b$) denotes the signal (background) probability density 
function (PDF), which is parameterized as a CB (ARGUS)-function with 
shape 
parameters determined from fits to MC samples and $N_s$, $N_b$, and 
$N_{p}$ are the  numbers of signal, 
combinatorial background and 
peaking background respectively. 
$N_s$ and $N_b$ are free parameters of the fit, while $N_{p}$ is 
fixed to the value obtained from fits to MC samples and scaled to the 
data luminosity ($N_p$ is set to zero for the $\tau \to \pi \nu$ mode). 
The fits are performed both for 
the three decay chains separately and for all chains combined
with a constraint to a common value of $\mathcal{B}(B^0\to 
D^{*-}\tau^+\nu_{\tau})$.
The fit results are included in Table~\ref{table-yields}. 
The total number of signal events is $60^{+12}_{-11}$ 
with a statistical significance of 6.7$\sigma$. The 
significance is defined as $\Sigma = 
\sqrt{-2{\ln}(\mathcal{L}_{\rm 0}/\mathcal{L}_{\rm max})}$, 
where $\mathcal{L}_{\rm max}$ 
and $\mathcal{L}_{\rm 0}$ denote the maximum likelihood value 
and the likelihood value for the zero signal hypothesis.
The fitted signal yield is used to calculate the branching fraction for 
the decay $B^0\to D^{*-}\tau^+\nu_{\tau}$ using the following formula,
which assumes equal fractions of charged and 
neutral $B$ mesons produced in $\Upsilon(4S)$ decays:
$\mathcal{B} = N_s/(N_{B\bar{B}}\times \sum_{ij}\epsilon_{ij}B_{ij})$,
where $N_{B\bar{B}}$ is the number of 
$B\bar{B}$ pairs, $\epsilon_{ij}$ 
denotes the reconstruction 
efficiency of the  specific decay chain and  $B_{ij}$ is the product 
of intermediate branching 
fractions  $\mathcal{B}(D^{*-}\to \bar{D}^0 
\pi^-)\times\mathcal{B}(\bar{D}^0\to i 
)\times\mathcal{B}(\tau^+ \to j)$. 
All the intermediate branching fractions are set to the 
PDG values \cite{PDG}. 
The branching fraction obtained
is $\mathcal{B}(B^0\to D^{*-}\tau^+\nu_{\tau} ) 
= (2.02^{+0.40}_{-0.37}(stat))$\%.

As a consistency check we also examine the distributions 
used in the signal selection, applying all requirements except 
those that are related to the considered variable. In all cases the 
distributions are well reproduced by the sum of signal and 
background components with normalizations fixed from the fits to 
the $M_{\rm tag}$ distribution.
We also  use the $M_{\rm mis}^2$ 
and $\cos\theta_{\nu_1\nu_2}$ 
(for $\tau \to \pi \nu$ mode) variables to extract the signal 
yield. 
We perform fits to 
distributions of these variables in the 
region $M_{\rm tag}>5.27$ GeV/$c^2$  
and obtain branching fractions 
in the range 1.83\% - 2.05\%
and in agreement
with the results 
from the 
$M_{\rm tag}$ fit.
   
\begin{table*}[t]
\caption{The number of expected combinatorial ($N_b^{\rm MC}$) and peaking 
($N_p$) background events, 
number of signal ($N_s$) and 
combinatorial background ($N_b$) 
events determined by 
the fits,
number of events in data ($N_{obs}$), 
signal selection efficiencies ($\epsilon$),
the product of the intermediate branching fractions (B),
extracted branching 
fraction
for $B^0\to D^{*-}\tau^+\nu_{\tau}$ ($\mathcal{B}$), statistical 
significance ($\Sigma$) and signal purity 
$S \equiv N_s/(N_s+N_b+N_p)$ in the 
$M_{\rm tag}>$5.27 GeV/$c^2$ region.
$N_s$, $\epsilon$ and B in the $\tau \to \pi \nu$ mode include cross 
feed events. 
The listed errors are statistical only.}
\begin{tabular}{ l r r c r r c r r c c c}
\hline\hline
subchannel & $N_{b}^{\rm MC}$&$N_p~~~$& $N_s$~~~~&$N_b$~~~ 
&$N_{obs}$&$ 
\epsilon 
\times 
10^{-4}$& 
B$\times 10^{-3}$&
$\mathcal{B}(\%)$~~&$\Sigma$&$S$\\
\hline 
$D^0\to K^-\pi^+$,$\tau \to e\bar{\nu}_e\nu_{\tau}$ & 
$26.3^{+5.4}_{-3.7}$& $1.2^{+1.6}_{-1.5}$ & $19.5^{+5.8}_{-5.0}$ & 
$~19.4^{+5.8}_{-5.0}~$
&40
& $3.25\pm 0.11$& ~4.59~~~~&
$2.44_{-0.65}^{+0.74}$&5.0$\sigma$&0.79\\

$D^0\to K^-\pi^+\pi^0$, $\tau \to e\bar{\nu}_e\nu_{\tau}$ 
& $50.8^{+5.5}_{-5.1}$&$5.0^{+2.6}_{-2.2}$ 
&$11.9^{+6.0}_{-5.2}$ &~$43.1^{+8.0}_{-7.2}~$ 
&60
&$0.78\pm 0.07$ &
17.03~~~~& 
$1.69^{+0.84}_{-0.74}$&2.6$\sigma$&0.50 \\

$D^0\to K^-\pi^+$,$\tau \to\pi^-\nu_{\tau}$ 
& $138.0^{+9.2}_{-8.8} $ &$-1.0^{+3.6}_{-3.2}$& 
$29.9^{+10.0}_{-~9.1}$&$118.0^{+14.0}_{-13.0}$
&148
&$1.07^{+0.17}_{-0.15}~~$&25.72~~~~& 
$2.02^{+0.68}_{-0.61}$&3.8$\sigma$&0.48 \\
\hline
Combined &
$215^{+12}_{-11}$&$6.2^{+4.7}_{-4.2}$ &$60^{+12}_{-11}$ &
$182^{+15}_{-14}~$
&248
&$1.17^{+0.10}_{-0.08}~~$& 47.34~~~~&
$2.02^{+0.40}_{-0.37}$&6.7$\sigma$&0.57 
\\
\hline\hline
\end{tabular}
\label{table-yields}
\end{table*}
We consider the following sources of systematic uncertainties in the 
branching
fraction determination.
The systematic error on $N_{B\bar{B}}$ is 1.3\%. 
The systematic uncertainties in 
the signal yield arise from uncertainties in the 
signal and background shape and peaking background. 
The systematic error 
due to the 
statistical uncertainties in the CB shape is  2.8\%. The 
CB parameters obtained from MC-samples are, within statistical 
errors, consistent with those extracted from fits to the control sample 
in data. Therefore we do not 
introduce additional uncertainties due to imperfect signal shape 
modeling. 
The systematic errors due to the parameterization of the 
combinatorial background are evaluated  by changing the ARGUS-shape 
parameters by $\pm 1\sigma$. 
Fits with the shape parameters 
allowed to float provide consistent results within statistical 
uncertainties.
The total systematic uncertainty due to 
the combinatorial background parameterization is  $^{+5.7}_{-10.1}$\%.  
The systematic error due to the peaking background is evaluated for each 
channel and amounts to $^{+8.2}_{-4.4}$\%  
for combined modes,
which is dominated by MC statistics. 
The effective efficiency  
$\sum_{ij}\epsilon_{ij}B_{ij}$ 
includes uncertainties in determination of the efficiencies for 
$B_{\rm tag}$ reconstruction,
($D^{*-}e^+/\pi^+$) pair 
selection and signal selection.
The uncertainty in  $B_{\rm tag}$  reconstruction is taken 
as the statistical error  in the $B_{\rm tag}$ efficiency 
evaluated from the data control sample (tagged with $B^0\to D^{*-}\pi^+$ 
decay) and is 10.9\%. The 
systematic error on the determination of  ($D^{*-}e^+/\pi^+$) pair 
selection efficiency comes from systematic uncertainties in the tracking 
efficiency, 
neutral reconstruction efficiency and particle identification and is in 
the range 7.9\%-10.7\% depending on the decay chain. 
Systematic uncertainties in the signal selection efficiency are 
determined by comparing MC and data distributions in the 
variables used for signal selection. 
The uncertainties due to the partial branching ratios 
are taken from the errors quoted in the PDG \cite{PDG}.  All of the 
above
sources of systematic uncertainties are combined together taking into 
account correlations between different decay chains.  The combined 
systematic uncertainty is 18.5\%.

We include the effect of systematic uncertainty in the signal 
yield on the significance of the observed signal by convolving the 
likelihood function from the fit  with a Gaussian systematic error 
distribution. The significance of the observed signal after including 
systematic uncertainties is 5.2$\sigma$.

In conclusion, in a sample of  
535$\times10^6~B\bar{B}$ pairs 
we observe a signal of 60$^{+12}_{-11}$  events for the decay 
$B^0\to D^{*-}\tau^+\nu_{\tau}$ with a significance of 5.2.
This is the first observation of an exclusive $B$ decay with 
the $b\to c \tau \nu_{\tau}$ transition.
The measured branching fraction: 
$\mathcal{B}(B^0\to D^{*-}\tau^+\nu_{\tau}) = (2.02^{+0.40}_{-0.37} 
(stat) \pm 0.37(syst))$\%
is consistent  within experimental uncertainties with SM 
expectations \cite{hwang}.

We thank the KEKB group for excellent operation of the
accelerator, the KEK cryogenics group for efficient solenoid
operations, and the KEK computer group and
the NII for valuable computing and Super-SINET network
support.  We acknowledge support from MEXT and JSPS (Japan);
ARC and DEST (Australia); NSFC and KIP of CAS (China); 
DST (India); MOEHRD, KOSEF and KRF (Korea); 
KBN (Poland); MES and RFAAE (Russia); ARRS (Slovenia); SNSF (Switzerland); 
NSC and MOE (Taiwan); and DOE (USA).



\begin{thebibliography}{99}

\bibitem{Itoh} H.~Itoh, S.~Komine and Y.~Okada, Progr. Theor. Phys. 
{\bf 114}, 179 (2005) and references quoted therein.
\bibitem{lep}
G.~Abbiendi {\it et al.} (OPAL Collaboration), Phys. Lett. B {\bf 520}, 
1 (2001);
R.~Barate {\it et al.} (ALEPH Collaboration), Eur. Phys. J. C  {\bf 19}, 
213
(1996);
P. Abreu {\it et al.} (DELPHI Collaboration), Phys. Lett. B {\bf 496}, 
43 
(2000);
M.~Acciarri {\it et al.} (L3 Collaboration), Z. Phys. C {\bf 71}, 379 
(1996).
\bibitem{PDG}
W.-M. Yao {\it et al.} (Particle Data Group), J. Phys. G {\bf 33}, 1 
(2006).
\bibitem{hwang}
J.~G.~K\"orner and G.~A.~Schuler, Phys. Lett. B {\bf 231}, 306 (1989); 
D.~S.~Hwang and D.~W.~Kim, Eur. Phys. J. C 
{\bf 14}, 271 (2000); 
C.-H.~Chen and C.-Q.~Geng, Phys. Rev. D {\bf 71}, 077501 (2005).  

\bibitem{CC}
Throughout this paper, 
the inclusion of the charge conjugate mode decay is implied
unless otherwise stated.

\bibitem{KEKB}
S.~Kurokawa and E.~Kikutani, Nucl. Instr. and. Meth. A {\bf 499}, 1 
(2003),
and other papers included in this volume.

\bibitem{Belle}
A.~Abashian {\it et al.} (Belle Collaboration),
Nucl. Instr. and Meth. A {\bf 479}, 117 (2002).

\bibitem{evtgen} 
D.~J.~Lange, 
Nucl. Instr. and Meth. A {\bf 462}, 152 (2001).
\bibitem{isgw2}D.~Scora and N.~Isgur, Phys. Rev. D {\bf 52}, 
2783 
(1995).
\bibitem{photos}E.~Barberio, Z.~W\c{a}s, Comput. Phys. Commun. {\bf 79}, 
291 
(1994).

\bibitem{Ikado}
K.~Ikado {\it et al.} (Belle Collaboration),
Phys. Rev. Lett. {\bf 97}, 061804 (2006).

\bibitem{CB}
T.~Skwarnicki, Ph.D.~Thesis, Institute of Nuclear Physics, Krakow 1986; 
DESY Internal Report, DESY F31-86-02 (1986).
\bibitem{ARGUS}H.~Albrecht et al. (ARGUS Collaboration), Phys. Lett. B 
{\bf 241}, 278 (1990).
\end{thebibliography}
\end{document}